\documentclass[10 pt,preprint]{aastex}
\usepackage {graphicx}
\slugcomment {Submitted to ApJ(Letters).}
\shortauthors {Lamb and Miller}
\shorttitle {Sonic-Point and Spin Resonance Model of Kilohertz QPOs}

\begin{document}

\title {Sonic-Point and Spin-Resonance Model of the Kilohertz QPO Pairs}

\author{Frederick K. Lamb\altaffilmark{1}}
\affil{Center for Theoretical Astrophysics, University of Illinois at Urbana-Champaign, 1110 W. Green St., Urbana, IL 61801-3080}
\email{fkl@uiuc.edu}

\and

\author{M. Coleman Miller}
\affil{Department of Astronomy, University of Maryland, College Park, MD 20742-2421}
\email{miller@astro.umd.edu}

\altaffiltext{1}{Also, Departments of Physics and Astronomy, University of Illinois at Urbana-Champaign.}

\begin {abstract}

Kilohertz quasi-periodic X-ray brightness oscillations (kilohertz QPOs) have now been detected in more than twenty accreting neutron stars in low-mass binary systems. Two kilohertz QPOs are usually detected in each star. Burst oscillations and two kilohertz QPOs have recently been detected in the 401~Hz accretion-powered X-ray pulsar \mbox{SAX~J1808.4$-$3658}. In this star the frequency of the burst oscillation is approximately equal to the star's spin frequency $\nu_{\rm spin}$ whereas the frequency separation of the two kilohertz QPOs is approximately $\nu_{\rm spin}/2$. If as expected the frequency of the burst oscillations in other stars is also approximately $\nu_{\rm spin}$, the frequency separation is $\approx \nu_{\rm spin}$ in some stars but $\approx \nu_{\rm spin}/2$ in others. A frequency separation approximately equal to $\nu_{\rm spin}/2$ is unexplained in all existing models of the kilohertz QPOs. Here we propose a modified version of the sonic-point beat-frequency model that can explain within a single framework why the frequency separation is close to $\nu_{\rm spin}$ in some stars but close to $\nu_{\rm spin}/2$ in others. As in the original sonic-point model, the frequency $\nu_{\rm QPO2}$ of the upper kilohertz QPO is close to the orbital frequency $\nu_{\rm orb}$ at the radius $r_{sp}$ of the sonic point in the disk flow. We show that magnetic and radiation fields rotating with the star will preferentially excite vertical motions in the disk at the ``spin-resonance'' radius $r_{sr}$ where $\nu_{\rm orb} - \nu_{\rm spin}$ is equal to the vertical epicyclic frequency. If the flow at $r_{sr}$ is relatively smooth, the vertical motions excited at $r_{sr}$ modulate the X-ray flux at $\nu_{\rm QPO1} \approx \nu_{\rm QPO2} - \nu_{\rm spin}$. If instead the gas at $r_{sr}$ is highly clumped, the vertical motions excited at $r_{sr}$ modulate the X-ray flux at $\nu'_{\rm QPO1} \approx \nu_{\rm QPO2} - \nu_{\rm spin}/2$. This \textit {sonic-point and spin-resonance} model can also explain quantitatively the decrease of the kilohertz QPO frequency separation with increasing accretion rate that is observed in many sources.

\end {abstract}


\section {Introduction}

Recent discoveries made with \textit {RXTE} have clarified the relationship between the pairs of kilohertz quasi-periodic oscillations (kilohertz QPOs) observed in the accretion-powered X-ray emission of neutron stars in low-mass binary systems (see van der Klis 2000; Lamb 2002), the high-frequency oscillations seen during thermonuclear X-ray bursts (see Strohmayer 2001; Strohmayer \& Bildsten 2003), and the spin frequency of the star. Key developments include the discovery of burst oscillations and kilohertz QPOs in the 401~Hz accretion-powered X-ray pulsar \mbox{SAX~J1808.4$-$3658} (Chakrabarty et al.\ 2003; Wijnands et al.\ 2003) and burst oscillations in the 314~Hz accretion-powered X-ray pulsar \mbox{XTE~J1814$-$338} (Markwardt \& Swank 2003; Markwardt, Strohmayer, \& Swank 2003b). 

\mbox{SAX~J1808.4$-$3658} was discovered and identified as an X-ray burst source during its September 1996 outburst (in 't Zand et al.\ 1998). During its April 1998 outburst, a 401~Hz periodic oscillation (``pulsation'') was detected in its persistent emission (Wijnands \& van der Klis 1998; Chakrabarty \& Morgan 1998), but neither kilohertz QPOs nor burst oscillations were detected then. Subsequent analysis of an X-ray burst recorded by BeppoSAX during the September 1996 outburst yielded a marginal detection of an oscillation at $400\pm2$~Hz (in 't Zand et al.\ 2001). \mbox{SAX~J1808.4$-$3658} was extensively observed throughout its October 2002 outburst and periodic oscillations, burst oscillations, and kilohertz QPOs were all detected with high confidence (Chakrabarty et al.\ 2003; Wijnands et al.\ 2003), making \mbox{SAX~J1808.4$-$3658} a Rosetta stone for understanding these phenomena.

The results of Chakrabarty et al.\ (2003) establish conclusively that the spin frequency of the neutron star in the \mbox{SAX~J1808.4$-$3658} system is 401~Hz: no oscillation at half this frequency was detected, and the upper limit on the rms amplitude of any such oscillation was determined to be 0.014\%, 300 times smaller than the amplitude of the 401~Hz oscillation. This excludes the possibility that the spin frequency is 200.5~Hz.

Wijnands et al.\ (2003) discovered a pair of kilohertz QPOs in \mbox{SAX~J1808.4$-$3658}. Their frequency separation $\Delta\nu_{\rm QPO}$ is within 3\% of $\nu_{\rm spin}/2$, demonstrating that $\Delta\nu_{\rm QPO}$ is related to the spin of the star and that the system somehow generates a frequency difference approximately equal to half the spin frequency, a result unexplained by existing models.

Chakrabarty et al.\ (2003) observed four X-ray bursts during the October 2002 outburst of\break\mbox{SAX~J1808.4$-$3658}. All showed a $\sim\,$401~Hz oscillation, demonstrating that in this star, $\nu_{\rm burst} \approx \nu_{\rm spin}$. Markwardt et al.\ (2003b) have analyzed three X-ray bursts from \mbox{XTE~J1814$-$338}. All showed nearly coherent $\sim\,$314~Hz oscillations, demonstrating that $\nu_{\rm burst} \approx \nu_{\rm spin}$ in this star, also. These results make compelling the previously strong evidence (Strohmayer et al. 1996, 1997, 1998a, 1998b; Strohmayer \& Markwardt 1999; Muno et al.\ 2000; Muno et al.\ 2002; Strohmayer \& Markwardt 2002) that burst oscillations are generated by the spin of the star and suggest that $\nu_{\rm burst}$ is close to $\nu_{\rm spin}$ in all the neutron stars in which burst oscillations have been observed. If so, then $\Delta\nu_{\rm QPO}$ is $\approx \nu_{\rm spin}$ in the stars with $\nu_{\rm spin} \lesssim 360$~Hz but is $\approx \nu_{\rm spin}/2$ in the stars with $\nu_{\rm spin} \gtrsim 400$~Hz. A successful model must therefore explain why some stars have $\Delta\nu_{\rm QPO} \approx \nu_{\rm spin}$, whereas others have $\Delta\nu_{\rm QPO} \approx \nu_{\rm spin}/2$.

Here we propose a modification of the original sonic-point beat-frequency model (Miller, Lamb, \& Psaltis, hereafter MLP98; Lamb \& Miller 2001) that can explain within a single framework why $\Delta\nu_{\rm QPO}$ is close to $\nu_{\rm spin}$ for some stars but close to $\nu_{\rm spin}/2$ for others. In \S~2 we discuss the general implications of the new observations for kilohertz QPO mechanisms and in \S~3 we propose specific mechanisms for generating the upper and lower kilohertz QPOs. We discuss our conclusions in \S~4.

\section {General Inferences from the Observations}

Taken together, the new and previous observations provide strong hints about the mechanisms that generate the kilohertz QPO pairs:

1.~~It appears highly likely that the frequency of one of the two kilohertz QPOs reflects the orbital frequency of gas in the inner disk. The kilohertz QPOs have frequencies similar to those of orbital motion near neutron stars. The frequencies of the kilohertz QPOs vary by hundreds of Hertz on time scales as short as minutes (see, e.g., M\'endez et al.\ 1999; van der Klis 2000). This is possible if their frequencies are related to orbital motion at a radius that varies (Lamb~2002). These frequency variations exclude mechanisms like the one proposed by Abramowicz \& Kluzniak (2001) to explain the high-frequency QPOs observed in black hole candidates and by Kluzniak et al.\ (2003) to explain the kilohertz QPOs observed in \mbox{SAX~J1808.4$-$3658}, because this type of mechanism requires a low-order resonance between the geodesic frequencies of test particles orbiting at a fixed radius and the resonance disappears when the two frequencies change.

2.~~The star's spin is somehow involved in producing the frequency separation of the two kilohertz QPOs in a pair. This involvement is clear in \mbox{SAX~J1808.4$-$3658}, where $\Delta\nu_{\rm QPO} \approx \nu_{\rm spin}/2$. It is strongly indicated in the other kilohertz QPO sources because in all cases where both $\Delta\nu_{\rm QPO}$ and $\nu_{\rm burst}$ have been measured, $\Delta\nu_{\rm QPO}$ is approximately commensurable with the stellar spin frequency inferred from $\nu_{\rm burst}$ (see van der Klis 2000; Lamb~2002). In four cases, $\Delta\nu_{\rm QPO}$ is consistent with $\nu_{\rm spin}$ or with half $\nu_{\rm spin}$ to better than 5\% (\mbox{4U~1608$-$52}: $\nu_{\rm spin}=319$~Hz, $\Delta\nu_{\rm QPO}=301.3\pm7.9$~Hz [M{\'e}ndez et al.\ 1997; M{\'e}ndez 2000, private communication]; \mbox{4U~1702$-$429}: $\nu_{\rm spin}=329$~Hz, $\Delta\nu_{\rm QPO}=333\pm5$~Hz [Markwardt, Strohmayer, \& Swank 1999]; \mbox{KS~1731$-$260}: $\nu_{\rm spin}=524$~Hz, $\Delta\nu_{\rm QPO}=260\pm 10$~Hz [Smith, Morgan, \& Bradt 1997; Wijnands \& van der Klis 1997]; and \mbox{SAX~J1808.4$-$3658}: $\nu_{\rm spin}=401$~Hz, $\Delta\nu_{\rm QPO}=195\pm6$~Hz [Chakrabarty et al.\ 2003; Wijnands et al.\ 2003]). In all other known cases, the largest value of $\Delta\nu_{\rm QPO}$ is approximately consistent with either $\nu_{\rm spin}$ or $\nu_{\rm spin}/2$.

3.~~A mechanism that produces a single sideband is indicated. Most mechanisms that modulate the X-ray brightness at two frequencies (such as amplitude modulation) would generate at least two strong sidebands. Although weak single and double sidebands have been detected close to the frequency of the lower kilohertz QPO (Jonker, M\'endez, \& van der Klis 2000), at most two strong kilohertz QPOs are observed in a given system (van der Klis 2000; M\'endez \& van der Klis 2000). This suggests that the frequency of one QPO is the primary frequency while the other is generated by a single-sideband mechanism. Beat-frequency mechanisms naturally produce a single sideband. Because one QPO frequency is almost certainly an orbital frequency, the most natural mechanism would be one in which the second frequency is generated by a beat with the star's spin frequency or with another orbital frequency. 

4.~~A successful model should explain why stars with high spin frequencies (as inferred from their burst oscillations) tend to have $\Delta\nu_{\rm QPO} \approx \nu_{\rm spin}/2$, whereas stars with low spin frequencies tend to have $\Delta\nu_{\rm QPO} \approx \nu_{\rm spin}$.

All existing beat-frequency models predict that if the fundamental frequencies are (1)~an orbital frequency $\nu_{\rm orb}$ and (2)~the frequency produced by a pattern with $n$-fold symmetry rotating with the star, then the principal beat frequency will be $n(\nu_{\rm orb}-\nu_{\rm spin})$ (Lamb et al.\ 1985). The sonic-point beat-frequency model does predict the appearance of weaker QPOs at a variety of other frequencies (MLP98). One of these could by chance differ from the fundamental orbital frequency by about $\nu_{\rm spin}/2$, but this would be an ad hoc explanation and would not explain a strong QPO at this frequency. Consequently, when evidence arose that in many cases $\nu_{\rm QPO1} \approx \nu_{\rm QPO2} - \nu_{\rm burst}/2$, it was suggested that perhaps in these stars $\nu_{\rm burst} \approx 2\nu_{\rm spin}$ (MLP98; see also Strohmayer et al.\ 1998b, Muno et al.\ 2001). The compelling evidence that $\nu_{\rm QPO1} \approx \nu_{\rm QPO2} - \nu_{\rm spin}/2$ in \mbox{SAX~J1808.4$-$3658} requires a model in which this frequency arises naturally. At the same time, beat-frequency models have many advantages, as discussed above. We propose a beat-frequency model in which $\nu_{\rm QPO1}$ is either approximately $\nu_{\rm QPO2} - \nu_{\rm spin}$ or approximately $\nu_{\rm QPO2} - \nu_{\rm spin}/2$.

\section {The Sonic-Point and Spin-Resonance Model}

We first show that the star's magnetic and radiation fields will excite vertical motion in the disk at the ``spin-resonance'' radius where $\nu_{\rm orb} - \nu_{\rm spin}$ equals the vertical epicyclic frequency $\nu_\psi$, and that the orbital frequency at this radius is approximately $\nu_{\rm spin}/2$. We propose that the upper kilohertz QPO is generated by clumps orbiting at the sonic point via the same mechanism as in the original sonic-point beat-frequency model (MLP98). Indeed, there is new evidence supporting this identification. We then show that interaction of the radiation pattern generated by clumps orbiting at the sonic-point radius $r_{sp}$ with the vertical motion of gas at the spin-resonance radius can produce a second QPO at either $\nu_{\rm orb}(r_{sp})-\nu_{\rm spin}/2$ or $\nu_{\rm orb}(r_{sp})-\nu_{\rm spin}$. We propose that this second QPO is the lower kilohertz QPO.

\subsection {Generation of motions with frequency $\nu_{\rm spin}/2$}

The magnetic and radiation fields of a neutron star are not perfectly symmetric about its rotation axis. Consequently, as the star spins and the gas in the disk orbits, each element of gas experiences periodically varying radial and vertical forces. If the frequency of this forcing is close to a natural frequency, the response of the element of gas may be large. The two most basic natural frequencies of an element of gas---aside from its orbital frequency---are its radial and vertical epicyclic frequencies. Vertical epicyclic motion appears more likely to generate a QPO, for two reasons. First, oscillation of an element of gas in the vertical direction can generate an X-ray signal by, for example, scattering radiation produced elsewhere in the system into or out of the observer's line of sight. In contrast, oscillation of an element of gas in the radial direction does not offer such possibilities. Second, vertical motion of gas in an accretion disk may be weakly damped  (Markovi{\'c} \& Lamb 1998) whereas radial motion is strongly damped (Markovi{\'c} \& Lamb 2000).

\begin{figure}[t]
\begin{center}
\includegraphics[angle=0, width=3.0 truein]{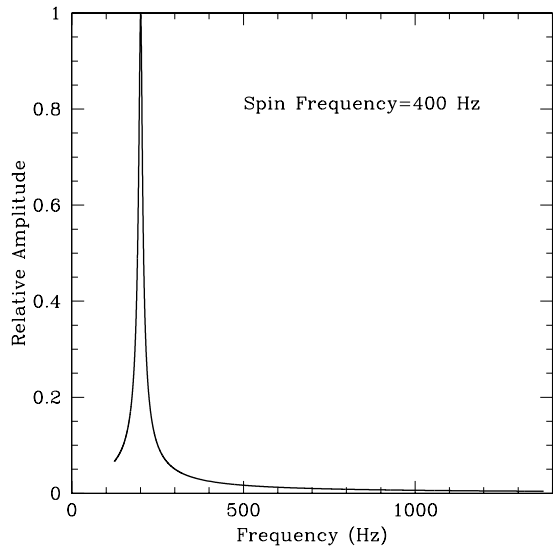}
\end{center}
\vspace{-20 pt}
\caption {Simulated response of gas in the accretion disk to a periodic perturbing force with frequency $\nu_{\rm spin}=400$~Hz. The vertical oscillation of the gas at the resonance radius has frequency $\nu_{\rm spin}/2$ and a much larger amplitude than at any other radius in the disk. In this simulation the damping time was assumed to be 1,000~s.}
\end{figure}

Perturbation of the disk flow by a magnetic or radiation field that rotates with the star will preferentially excite vertical motion in the disk at the spin-resonance radius $r_{sr}$ defined implicitly by the resonance condition 
\begin{equation}
\nu_{\rm spin}-\nu_{\rm orb}(r_{sr})=\nu_{\psi}(r_{sr})\;,
\end{equation}
where $\nu_\psi(r)$ is the vertical epicyclic frequency at radius $r$. Here $\nu_{\rm spin}-\nu_{\rm orb}(r_{sr})$ is the forcing frequency experienced by an element of gas orbiting at $r_{sr}$. Numerical simulations (see Fig.~1) show that the vertical displacement of gas is much greater at the resonant radius than at any other radius. In a Newtonian $1/r$ gravitational potential, $\nu_\psi(r)=\nu_{\rm orb}(r)$. In general relativity, $\nu_\psi(r)$ is not exactly equal to $\nu_{\rm orb}(r)$, but the difference is $<2$~Hz at the radii of interest (where $\nu_{\rm orb}<300$~Hz). Consequently, at the resonance radius where vertical motion is preferentially excited, $\nu_{\rm orb} \approx \nu_\psi \approx \nu_{\rm spin}/2$, i.e., at this radius the orbital and vertical frequencies are both approximately half the star's spin frequency.

Each \textit {individual} element of gas at the spin-resonance radius orbits the star with frequency $\nu_{\rm orb}(r_{\rm sr}) \approx \nu_{\rm spin}/2$. However, the \textit {pattern} created by the elements of gas that are above the plane moves around the star with frequency $\nu_{\rm pattern} = \nu_{\rm spin}$. Stated differently, the azimuth where the gas is highest above the disk advances at the rate $\nu_{\rm spin}$, even though no gas moves at this rate. The reason for this is that the response of the gas is almost in phase with the periodic driving force. For example, if the force on a particular element of gas is maximal at the time-dependent azimuthal position $\phi_{\rm max}(t) = \phi_0 + \nu_{\rm spin}t$ and when the gas is furthest above the disk plane, then as the forcing field rotates, the gas in the disk at $r_{sr}$ will be at the peak of its vertical excursion at the azimuth $\phi=\phi_{\rm max}(t)$.

So far we have considered the effect of forces acting only on one (the `upper'') surface of the disk. The second harmonic of $\nu_{\rm spin}$ is weak in the persistent X-ray emission of all known millisecond accretion-powered pulsars and, where both have been determined, $\nu_{\rm burst}$ is close to $\nu_{\rm spin}$ rather than $2\nu_{\rm spin}$, indicating that we observe only one (the ``upper'') magnetic pole of the star and its X-ray beam as the star rotates. The star's opposite pole and its X-ray beam interact with clumps when they are below the lower surface of the disk with a phase that will further amplify their vertical motions.

\subsection {Generation of the upper kilohertz QPO}

We propose that the upper kilohertz QPO is related to the orbital motion of clumps at a special radius near the neutron star. As shown in MLP98, the radius $r_{sp}$ of the sonic point in the disk is a strong candidate for this special radius. The sonic transition at $r_{sp}$ is a strong-field general relativistic effect and is usually produced by the force on the gas in the disk exerted by radiation from the star. At $r_{sp}$, the inward radial velocity in the accretion disk increases sharply. The sharpness of this change in the radial velocity matches the observed narrow width of the upper kilohertz QPO in power density spectra. In addition, radiation and magnetic forces are likely to enhance clumping at the sonic radius over clumping further out in the disk. As in the original sonic-point model, we assume the frequency $\nu_{\rm QPO2}$ of the upper kilohertz QPO is generated by mapping of accretion streams from clumps orbiting at $r_{sp}$ onto the surface of the star. This mechanism for generating the upper kilohertz QPO is supported by the observed short-term anticorrelation of the frequency of the upper kilohertz QPO and the X-ray flux in \mbox{4U~1608$-$52} (Yu, van der Klis, \& Jonker 2001; Yu \& van der Klis 2002).

As discussed in Lamb \& Miller (2001), the inward drift of the clumps at $r_{sp}$ causes the footpoints of the streams on the stellar surface to revolve around the star with an angular velocity slightly different from $\nu_{\rm orb}(r_{sp})$. The radiation pattern they produce rotates with frequency $\nu_{\rm QPO2} \equiv \nu_{\rm orb}(r_{sp}) - \delta\nu_{\rm rad}$, where $\delta\nu_{\rm rad}$ is typically positive and increases in magnitude from nearly zero to a few tens of Hertz as $r_{sp}$ decreases.

\subsection {Generation of the lower kilohertz QPO}

Scattering of radiation from the neutron star by the clumps orbiting in the disk close to the sonic radius will create a radiation pattern on the surface of the disk that rotates with frequency $\nu_{\rm orb}(r_{\rm sp})$. This pattern will preferentially illuminate or shadow the gas orbiting at the spin-resonance radius, when it is above the disk plane. The rotating radiation pattern will be scattered much more strongly by clumps orbiting at the spin-resonance radius than by clumps elsewhere because clumps at this radius are making large vertical excursions, unlike clumps elsewhere in the disk.

\begin{figure*}[t]
\begin{minipage}[t]{3.2truein}
\includegraphics[angle=0, width=3.0 truein]{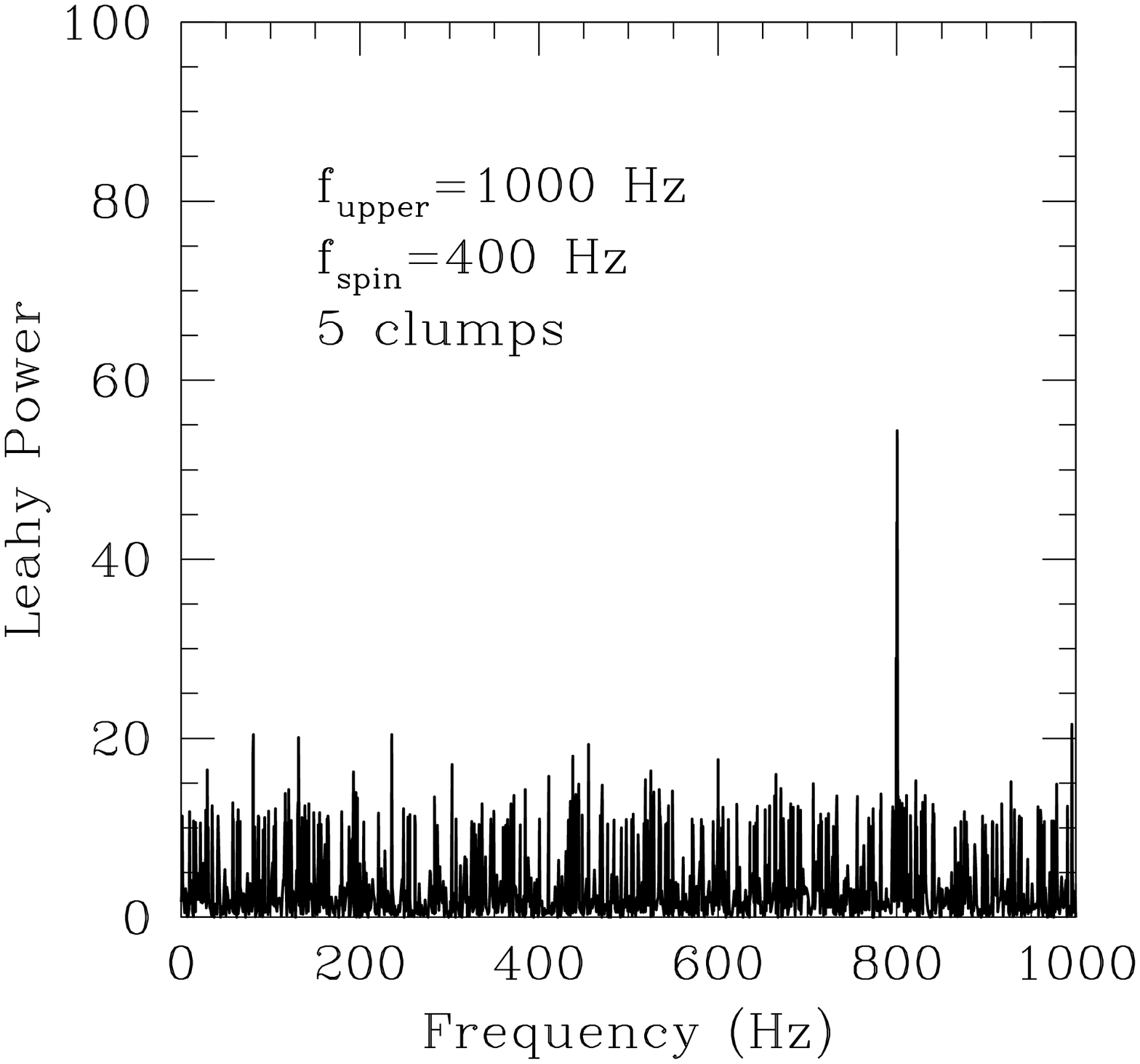}
\end{minipage}
\hfill
\begin{minipage}[t]{3.2truein}
\includegraphics[angle=0, width=3.0 truein]{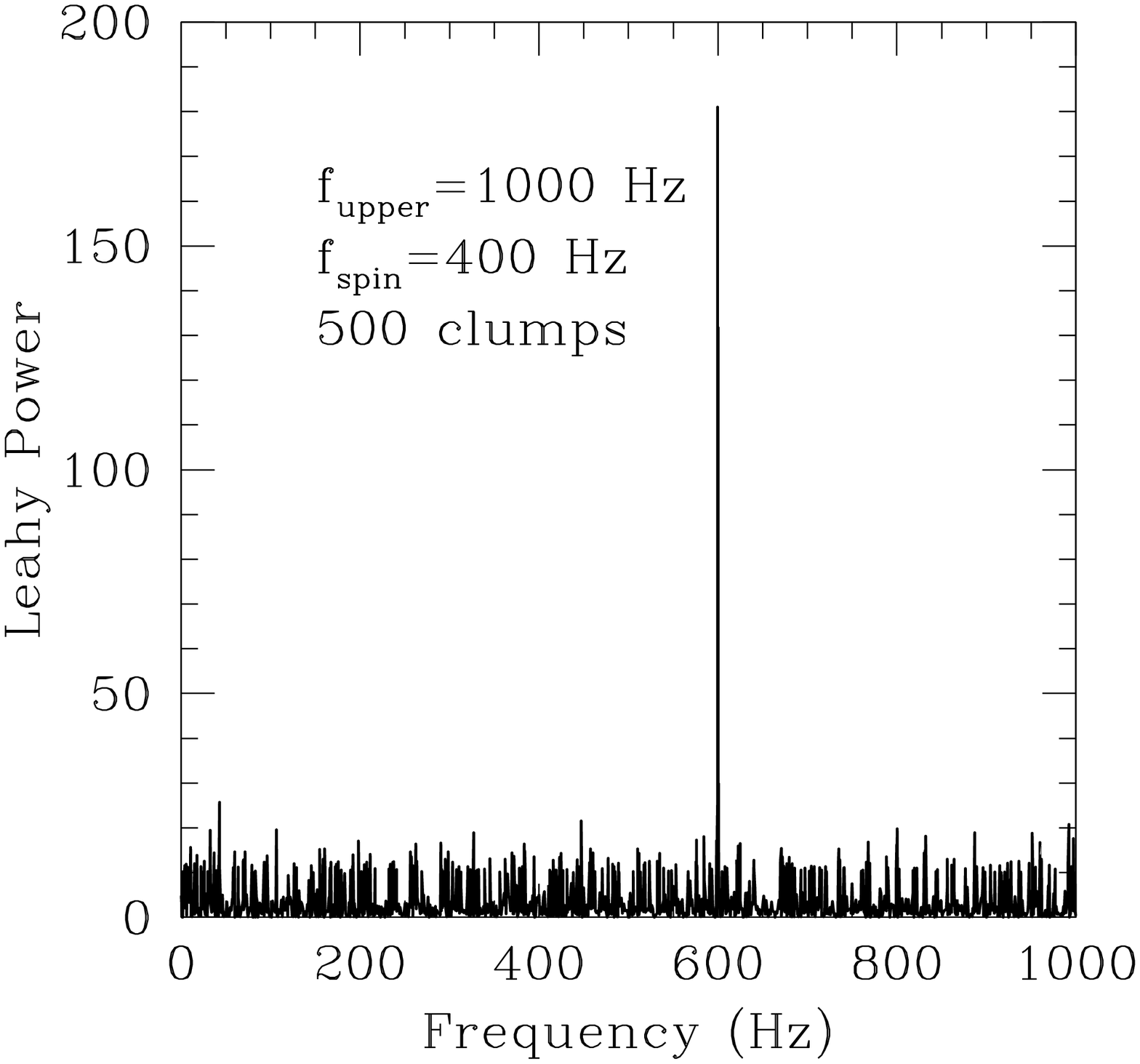}
\end{minipage}
\vspace{-15 pt}
\caption[fig1]{
\large\normalsize
Power spectra of the X-ray flux modulation produced by simulations of a disk with a small number of clumps near the spin-resonance radius (left-hand panel) and a large number of clumps (right-hand panel). The star's spin frequency is 400~Hz while the orbital frequency at the sonic point is 1,000~Hz. These power spectra demonstrate that if the flow near the spin-resonance radius is clumpy, the effect of individual clumps dominates and the dominant frequency is $\nu_{\rm orb}(r_{\rm sp}) - \nu_{\rm spin}/2$. If instead the flow is relatively smooth, the effect of the clump pattern dominates and the dominant frequency is $\nu_{\rm orb}(r_{\rm sp}) - \nu_{\rm spin}$. This simulation did not include any signal with the orbital frequency of the gas at the sonic radius.}
\end{figure*}

\textit {Generation of a lower kilohertz QPO with frequency $\nu_{\rm QPO1} \approx \nu_{\rm orb}(r_{sp}) - \nu_{\rm spin}/2$}.---Suppose first that the gas in the disk near the spin-resonance radius is highly clumped. When illuminated, each clump orbiting at $r_{sr}$ scatters radiation in all directions. In effect, each clump redirects the radiation propagating outward from the sonic radius in the modest solid angle that it subtends (as seen from the sonic radius) into all directions. From the point of view of a distant observer, each individual clump looks like a light bulb that is blinking on and off with a frequency equal to $\nu_{\rm orb}(r_{sp}) - \nu_{\rm orb}(r_{sr}) \approx \nu_{\rm orb}(r_{sp}) - \nu_{\rm spin}/2$. If there are only a modest number of clumps at $r_{sr}$, the effect is somewhat like what one would see if a movie marquee had only a few light bulbs, so the blinking of the individual bulbs dominates the time variation. Because the radiation is scattered in all directions, an observer does not have to be close to the disk plane to see the X-ray flux modulation. The left-hand panel of Figure~2 shows the power spectrum of the flux variation generated in a simulation in which five randomly-positioned clumps scatter the radiation pattern coming from the sonic radius. The peak at $\nu_{\rm orb}(r_{\rm sp}) - \nu_{\rm spin}/2$ is clearly dominant.

\textit {Generation of a lower kilohertz QPO with frequency $\nu_{\rm QPO1} \approx \nu_{\rm orb}(r_{\rm sp}) - \nu_{\rm spin}$}.---Suppose now that the gas in the disk near the spin-resonance radius is less highly clumped. There may be a larger number of smaller clumps or the flow may even be relatively smooth. As before, each element of gas is oscillating vertically with frequency $\nu_{\rm spin}/2$. Together they form a pattern of raised fluid elements that rotates around the star with frequency $\nu_{\rm spin}$. Because a large number of fluid elements are scattering radiation to the observer at any given moment, their individual contributions blend together, so the dominant time variation has frequency $\nu_{\rm orb}(r_{sp}) - \nu_{\rm spin}$. The effect is somewhat like what one sees when viewing a movie marquee with hundreds of light bulbs blinking in phase so that the pattern formed by the lighted bulbs moves around the marquee. In this case the brightness variation produced by the pattern of lighted bulbs dominates the brightness variation produced by the individual bulbs. The right-hand panel of Figure~2 shows the power spectrum of the flux variation generated in a simulation in which 500 randomly-positioned clumps scatter the radiation pattern coming from the sonic radius. The peak at $\nu_{\rm orb}(r_{sp}) - \nu_{\rm spin}$ is clearly dominant.

\begin{table*}[t]
\begin{center}
\begin{tabular}{lcccc}
\hline
\noalign{\vspace{2 pt}}
\hline
\noalign{\vspace{3 pt}}
\multicolumn{5}{c}{\bfseries Table 1:
Fits to $\Delta\nu_{\rm QPO}$ vs. $\nu_{\rm QPO2}$ data}\\
\noalign{\vspace{3 pt}}
\hline
\noalign{\vspace{3 pt}}
Source&M($M_\odot$)&$\Delta\nu_0$(Hz)&coeff&$\chi^2$/dof\\
Sco X--1&1.4&365&0.0162&96.3/46\\
4U~1608--52&1.96&310&0.0006&22.1/10\\
4U~1728--34&1.67&364&0.0017&7.3/6\\
4U~1820--30&2.0&279&0.00001&7.2/18\\
\noalign{\vspace{3 pt}}
\hline
\noalign{\vspace{3 pt}}
\end{tabular}
\end{center}
\end{table*}

\textit {Variation of the kilohertz QPO frequency difference}.---In the model proposed here, the frequency of the upper kilohertz QPO is close to but less than $\nu_{\rm orb}(r_{sp})$ and the difference varies by as much as several tens of Hertz. However, the frequency of the lower kilohertz QPO is very close to either $\nu_{\rm orb}(r_{sp}) - \nu_{\rm spin}$ or $\nu_{\rm orb}(r_{sp}) - \nu_{\rm spin}/2$, depending on whether the flow at $r_{sp}$ is smooth or clumpy. Consequently, the frequency separation between the two QPOs is not exactly $\nu_{\rm spin}$ (or $\nu_{\rm spin}/2$). Table~1 shows the results of fits of this model to the observed changing frequency separation in several sources (see van der Klis et al.\ 1997; M\'endez et al.\ 1998a; M\'endez et al.\ 1998b; Ford et al.\ 1998; Psaltis et al.\ 1998; M\'endez \& van der Klis 1999; Jonker, M\'endez, \& van der Klis 2002), adopting the model of Lamb \& Miller (2001; note that in that work the frequency of the lower kilohertz QPO also varies by a few tens of Hertz, whereas here only the frequency of the upper kilohertz QPO varies). The fit is reasonable for all these sources except \mbox{Sco~X-1}, which has an unusually complicated $\Delta\nu_{\rm QPO}$--$\nu_{\rm QPO2}$ relation.

\section{Discussion and Conclusions}

The sonic-point and spin-resonance model proposed here appears able to explain the observed frequencies of the two kilohertz QPOs, the variation of their frequencies by hundreds of Hertz, the closeness of their frequency separation $\Delta\nu_{\rm QPO}$ to $\nu_{\rm spin}$ in some sources and to $\nu_{\rm spin}/2$ in others, and the variation of $\Delta\nu_{\rm QPO}$ with $\nu_{\rm QPO2}$ (accretion rate) in a given source.

\textit {The difference between ``fast'' and ``slow'' rotators}.---In the model proposed here, $\nu_{\rm QPO1} \approx \nu_{\rm orb}(r_{\rm sp}) - \nu_{\rm spin}/2$ if the resonance with the stellar spin occurs where the gas in the disk is highly clumped, whereas $\nu_{\rm QPO1} \approx \nu_{\rm orb}(r_{\rm sp}) - \nu_{\rm spin}$ if the resonance occurs where the gas in the disk is relatively smooth. Magnetic forces may cause the gas in the accretion disk to become more clumped as it approaches the neutron star (MLP98; Lamb \& Miller 2001). Consequently, the parameters that may be most important in determining whether the flow at the spin resonance radius $r_{sr}$ is clumpy or smooth are the star's spin frequency and magnetic field. For a given stellar magnetic field, the flow is likely to be more clumpy if the star is spinning rapidly and $r_{sr}$ is therefore close to the star. For a given spin rate, the flow is likely to be more clumpy if the star's magnetic field is stronger.

The four sources with $\nu_{\rm spin}>400$~Hz and measurable frequency separations have $\Delta\nu_{\rm QPO} \approx \nu_{\rm spin}/2$ whereas the three sources with $\nu_{\rm spin}<400$~Hz have $\Delta\nu_{\rm QPO} \approx \nu_{\rm spin}$. With such a small sample, one cannot make any definite statements, but the apparent trend is consistent with the sonic-point and spin-resonance model. The model suggests that if kilohertz QPOs are detected in the recently-discovered 185~Hz, 191~Hz, and 314~Hz accretion-powered X-ray pulsars \mbox{XTE~J0929$-$314} (Galloway et al.\ 2002), \mbox{XTE~J1807$-$294} (Markwardt, Smith, \& Swank 2003a), and \mbox{XTE~J1814$-$338} (Markwardt \& Swank 2003), their frequency separations should be approximately equal to the respective spin frequencies. The 435~Hz spin frequency of \mbox{XTE~J1751$-$305} (Markwardt et al.\ 2002) is high enough that $\Delta\nu_{\rm QPO}$ could be either approximately 435~Hz or approximately 217~Hz; QPOs at both frequencies might even be detectable.

\textit {Other oscillation frequencies}.---We expect the response at the spin-resonance radius defined by $\nu_{\rm orb}\approx\nu_{\rm spin}/2$ to be largest, but there may be other, weaker motions at other low-order resonances. In principle, weak beat-frequency QPOs could be generated by responses at the radii where $\nu_{\rm orb}/\nu_{\rm spin}$ is close to ratios of small integers, such as 1/3, 2/3, and so on. The strength of such secondary QPOs would depend on the smoothness of the flow and the rates at which vertical motions are damped at these radii. If the flow is clumpy, so that the primary beat is with the local orbital frequency, then several beat frequencies could be observed. If the flow is instead smooth, so that the primary beat is with the pattern frequency, then because the pattern frequency is always the spin frequency, no other significant QPOs would be expected. In either case, the stronger the damping is, the weaker the other QPOs would be relative to the one at the fundamental spin-resonance radius.

We note that there is no known reason why the mechanism for producing a lower kilohertz QPO proposed in the original sonic-point beat-frequency model would not operate. Apparently this mechanism does not produce a strong QPO in the fast rotators, but it might produce a weak QPO in these sources. If it operates in the slow rotators, it would produce a QPO near $\nu_{\rm orb}(r_{sp}) - \nu_{\rm spin}$ that might appear as a sideband to the lower kilohertz QPO.

Although the model outlined here is qualitatively consistent with the basic properties of the kilohertz QPOs, many aspects of the model require further exploration and development. For example, what if any sidebands or additional QPOs are to be expected? How large are the vertical excursions that would be expected in a model that includes more of the physics of accretion disks? Is the model quantitatively consistent with the observational upper limits on X-ray oscillations at $\nu_{\rm spin}$? It will be important to pursue these questions, but in the meantime resonant excitation of gas in the disk appears to be worth exploring as a mechanism for generating the lower kilohertz QPO.

\acknowledgements

We thank Deepto Chakrabarty and Michiel van der Klis for helpful discussions. This work was supported in part by NSF grant AST~0098399 and NASA grant NAG5-12030 at Illinois, and by NSF grant AST~0098436 at Maryland.

\end{document}